\documentstyle[prl,floats,aps,twocolumn,multicol,epsf]{revtex}
\tighten
\begin{document}
\twocolumn[\hsize\textwidth\columnwidth\hsize\csname
@twocolumnfalse\endcsname
\title{Dipole perturbations of the Reissner-Nordstr\"{o}m
solution: II. The axial case}
\author{Lior M. Burko}
\address{
Department of Physics,
Technion---Israel Institute of Technology, 32000 Haifa, Israel}
\date{\today}

\maketitle

\begin{abstract}
We study the linear metric perturbations of the
Reissner-Nordstr\"{o}m
solution for the case of axial perturbation modes. We find that the
well-known perturbative analysis fails for the case of dipole
$(l=1)$ perturbations, although valid for higher multipoles. We define new
radial functions, with which the perturbation formalism is
generalized to all multipole orders, 
including the dipole. We then complete the solution by
constructing the perturbed metric and Maxwell tensors.
\newline
\newline
PACS number(s): 04.70.Bw, 04.25.Nx
\end{abstract}

\vspace{3ex}
]

\section{Introduction}
The linear gravitational and electromagnetic perturbations of the 
Reissner-Nordstr\"{o}m solution have been analyzed by Moncrief 
\cite{moncrief}, Zerilli \cite{zerilli},  
Chandrasekhar 
\cite{Chandra1}, and Chandrasekhar and Xanthopoulos \cite{Chandra2}. (For a 
summary of these works see Refs. \cite{chandra,frolov}. These perturbations
were
considered also in \cite{mcnamara,matzner,gursel,chandra-hartle}.) In
the perturbative 
approach the deviations from the exact Reissner-Nordstr\"{o}m solution are
considered as (small) perturbations over the 
Reissner-Nordstr\"{o}m background. Due to the coupling of the 
electromagnetic and the gravitational fields in the background, any 
gravitational perturbation leads in general via the field equations to an 
electromagnetic perturbation, and vice versa. Despite this complication it 
turns out that the equations describing the evolution of the 
perturbations can be decoupled (for each multipole order and for each 
parity) to a pair of independent one-dimensional 
wave equations (namely, the Moncrief-Zerilli equations), each of which is 
made of an electromagnetic component and a gravitational component. 

In a recent paper \cite{burko} (hereafter Paper I) 
we analyzed the polar modes of the 
perturbations, and showed that the perturbative formalism of Ref. 
\cite{chandra} fails in the case of dipole perturbations (namely, 
the $l=1$ modes in a multipole expansion). We generalized the perturbative
analysis to include also the dipole modes by 
defining new radial functions, and completed the solution by giving 
the metric perturbations and the perturbations of the Maxwell tensor in 
terms of the perturbing fields. In this sequel we treat the axial 
perturbations of the Reissner-Nordstr\"{o}m solution. (For a definition of 
the parities see, e.g., Ref. \cite{chandra}.) As we shall see in what follows, 
also in the axial 
case the formalism of Ref. \cite{chandra} fails, and for similar reasons to 
the failure in the polar case discussed in Ref. \cite{burko}. In this paper 
we shall again define new radial functions, with which one can formulate a
perturbative analysis which is valid for all multipole orders, including
the dipole. 

For many applications the failure of the perturbative formalism for the 
dipole modes is immaterial. This is due to the non-radiative character of 
dipole gravitational modes. However, there are cases where dipole 
{\it electromagnetic} waves are important, especially in the
Reissner-Nordstr\"{o}m spacetime, because of the coupling of the
electromagnetic and gravitational fields. For example, the late-time
behavior of the electromagnetic perturbations produced during the collapse
is controlled by the dipole modes (which are the least-damped modes)
\cite{price,bicak,burko-ori97}. In addition, the dipole perturbations are
especially important in the analysis of the (electromagnetic)
perturbations of the Cauchy horizon of Reissner-Nordstr\"{o}m black holes 
\cite{burko-ori95}.  
One needs, therefore, a valid perturbative formalism for 
the calculation of the perturbed metric coefficients and the components of 
the Maxwell tensor. 

The organization of this paper is as follows: In section II we discuss our  
definitions and the notations. In section III we review the 
perturbative approach 
we use. In section IV we derive the coupled equations, 
which we decouple in section V, and in section VI we complete the 
solution, which is adequate for the treatment of the dipole modes. 

\section{Definitions and Notation}

Following the notation and convention of Ref. \cite{chandra}, we
write the line element of an unperturbed Reissner-Nordstr\"{o}m
black hole in the form 
\begin{eqnarray} 
 \,ds^{2}=e^{2\nu}\left(\,dx^{0}\right)^{2}-e^{2\mu _{2}}
\left(\,dx^{2}\right)^{2}
-r^{2}\,d\Omega^{2}, \end{eqnarray}
where $\,d\Omega^{2}=e^{2\psi}\left(\,dx^{1}\right)
^{2}+e^{2\mu _{3}}\left(
\,dx^{3}\right)^{2}$ is the line element on the  
unit two-sphere, the metric
coefficients are $e^{2\nu}=e^{-2\mu_{2}}=(r^{2}-2Mr+Q_{*}^{2})/r^{2}
\equiv \Delta/r^{2}$ and $e^{2\psi}=e^{2\mu_{3}}\sin^{2}\theta=
r^{2}\sin^{2}\theta$, and the coordinates are 
$\left(x^{0}\;x^{1}\;x^{2}\;x^{3}\right)=\left(t\;\phi\;r\;\theta\right).$ 
Here, $r$ is the radial Schwarzschild coordinate, defined such that 
circles of radius $r$ have circumference $2\pi r$, and $M$ and 
$Q_{*}$ are the mass and electric charge, correspondingly, of the black hole. 

It turns out that the metric perturbations of the Reissner-Nordstr\"{o}m 
black hole can be separated to two parities---called polar (or even 
parity) and axial (or odd parity)---in accordance with their behavior
under the transformation $\phi\rightarrow -\phi$. The Reissner-Nordstr\"{o}m 
background, being spherically symmetric and static, 
is characterized by the vanishing of all the axial metric coefficients 
[namely, in the background one has 
$\omega=q_{2}=q_{3}=0$; see the definitions for 
these metric coefficients below in Eq. (\ref{metric})]. Thus, any axial
perturbation will impart rotation on 
the black hole, and transform it into a (slowly) rotating Kerr-Newman black 
hole. It can be shown \cite{chandra} that the metric 
\begin{eqnarray}
\,ds^{2}&=&e^{2\nu}\left(\,dx^{0}\right)^{2}-e^{2\psi}\left(\,dx^{1}-\omega
\,dx^{0}-q_{2}\,dx^{2}\right.\nonumber\\
&-&\left.q_{3}\,dx^{3}\right)^{2}
-e^{2\mu_{2}}\left(\,dx^{2}\right)^{2}-e^{2\mu_{3}}
\left(\,dx^{3}\right)^{2}.
\label{metric}
\end{eqnarray} 
is of sufficient generality for the treatment of all perturbations (polar 
and axial). 

Throughout this paper we shall use the notation of Ref. \cite{chandra}, 
except when we change its formalism; Then, similarly to the notation of 
Paper I, we shall add a bar to the symbols of 
Ref. \cite{chandra}. The `barred' objects will be defined in such a way 
that dipole perturbations are treated adequately. This notation
facilitates the comparison between our formalism, and the formalism of
Ref. \cite{chandra}.

\section{The perturbative formalism}
In a similar way to the development of the perturbative equations for the 
polar modes, the equations governing the axial perturbations are derived by 
a linearization of the coupled Einstein-Maxwell equations about the 
Reissner-Nordstr\"{o}m background \cite{chandra}. In particular, after 
substitution of the expressions for the unperturbed metric coefficients (of 
the background) in the linearized $\phi \, r$ and $\phi \, \theta$
components of the Ricci tensor, one 
obtains the equations 
\begin{eqnarray}
\left[r^{2}e^{2\nu}(q_{2,3}-q_{3,2})\sin^{3}\theta\right]_{,3}&+& 
r^{4}(\omega_{,2}-q_{2,0})_{,0}\sin^{3}\theta\nonumber\\
&=&2r^{3}e^{\nu}\sin^{2}\theta \,\delta R_{(1)(2)}\nonumber\\
&=&4Q_{*}re^{\nu}F_{(0)(1)}\sin^{2}\theta
\label{ricci1}
\end{eqnarray}
and 
\begin{eqnarray}
\left[r^{2}e^{2\nu}(q_{2,3}-q_{3,2})\sin^{3}\theta\right]_{,2}&-&
r^{2}e^{-2\nu}(\omega_{,3}-q_{3,0})_{,0}\sin^{3}\theta\nonumber\\
&=&-2r^{2}\sin^{2}\theta \,\delta R_{(1)(3)}\nonumber\\
&=&0, 
\label{ricci2}
\end{eqnarray}
where $X_{(\alpha)(\beta)}$ is the $(\alpha)(\beta)$ tetrad component of 
the tensor $X$. Here, $R$ and $F$ are the Ricci and Maxwell tensors, 
respectively, and a comma denotes partial differentiation. Substitution in 
the linearized components of the Maxwell tensor yields 
\begin{eqnarray}
\left[re^{\nu}F_{(0)(1)}\right]_{,2}+re^{-\nu}
F_{(1)(2),0}=0,
\label{max1}
\end{eqnarray}
\begin{eqnarray}
re^{\nu}\left[F_{(0)(1)}\sin\theta\right]_{,3}+r^{2}F_{(1)(3),0}
\sin\theta=0,
\label{max2} 
\end{eqnarray}
\begin{eqnarray}
re^{-\nu}F_{(0)(1),0}&+&\left[re^{\nu}F_{(1)(2)}\right]_{,2}+F_{(1)(3),3}
\nonumber\\
&=&-Q_{*}(\omega_{2}-q_{2,0})\sin\theta.
\label{max3}
\end{eqnarray}
We now introduce the functions 
\begin{eqnarray}
B(r,\theta)=F_{(0)(1)}\sin\theta,
\label{b}
\end{eqnarray}
\begin{eqnarray}
Q(r,\theta)=r^{2}e^{2\nu}(q_{2,3}-q_{3,2})\sin^{3}\theta.
\label{q}
\end{eqnarray}
After substitution of these functions in Eqs. 
(\ref{ricci1}) and (\ref{ricci2}), and using the linearized Maxwell 
equations [Eqs. 
(\ref{max1})--(\ref{max3})], we obtain the differential equations
\begin{eqnarray}
\frac{1}{r^{4}\sin^{3}\theta}\frac{\,\partial Q}{\,\partial \theta}=
-(\omega_{,2}-q_{2,0})_{,0}+\frac{4Q_{*}}{r^{3}\sin^{2}\theta}
e^{\nu}B 
\label{dif1}
\end{eqnarray} 
and 
\begin{eqnarray}
\frac{\Delta}{r^{4}\sin^{3}\theta}\frac{\,\partial Q}{\,\partial r}=
(\omega_{,3}-q_{3,0})_{,0}.
\label{dif2}
\end{eqnarray}
We now assume that the perturbations can be analyzed to their normal modes 
with a time dependence 
$e^{i\sigma t}$. (This is always possible for a linear theory.) Because of
this explicit time-dependence of the metric perturbations we did not
consider the time dependence in the definitions for the functions
$B(r,\theta)$ and $Q(r,\theta)$. Substitution in Eqs. (\ref{dif1}) and
(\ref{dif2}) yields then \begin{eqnarray}
\frac{1}{r^{4}\sin^{3}\theta}\frac{\,\partial Q}{\,\partial \theta}=
-i\sigma\omega_{,2}-\sigma^{2}q_{2}+\frac{4Q_{*}}{r^{3}\sin^{2}\theta}
e^{\nu}B
\label{dif3}
\end{eqnarray}
and
\begin{eqnarray}
\frac{\Delta}{r^{4}\sin^{3}\theta}\frac{\,\partial Q}{\,\partial r}=
i\sigma\omega_{,3}+\sigma^{2}q_{3}.
\label{dif4}
\end{eqnarray}
Differentiating Eq. (\ref{dif3}) with respect to $\theta$ and Eq. 
(\ref{dif4}) with respect to $r$, and summing the equations, we find that 
\begin{eqnarray}
r^{4}\frac{\partial}{\,\partial r}\left(\frac{\Delta}{r^{2}}
\frac{\,\partial Q}{\,\partial r}\right)&+&\sin^{3}\theta
\frac{\partial}{\,\partial\theta}\left(\frac{1}{\sin^{3}\theta}
\frac{\,\partial Q}{\,\partial\theta}\right)+\sigma^{2}\frac{r^{4}}
{\Delta}Q\nonumber\\
&=&4Q_{*}e^{\nu}r\frac{\partial}{\,\partial\theta}
\left(\frac{B}{\sin^{2}\theta}\right)\sin^{3}\theta.
\label{eq1}
\end{eqnarray} 
We now differentiate Eq. (\ref{max1}) with respect to $r$, Eq. 
(\ref{max2}) with respect to $\theta$, and Eq. (\ref{max3}) with respect to 
$t$. Substituting in the latter and using Eq. (\ref{b}), we find  
\begin{eqnarray}
\left[e^{2\nu}(re^{\nu}B)_{,2}\right]_{,2}&+&\frac{e^{\nu}}{r}
\left(\frac{B_{,3}}{\sin\theta}\right)_{,3}\sin\theta-re^{-\nu}B_{,0,0}
\nonumber\\
&=&Q_{*}(\omega_{,2,0}-q_{2,0,0})\sin^{2}\theta.
\label{eq2}
\end{eqnarray}
Eqs. (\ref{eq1}) and (\ref{eq2}) govern the axial 
perturbations.

We next separate the variables in $B(r,\theta)$ and $Q(r,\theta)$. This 
separation is done in Ref. \cite{chandra} by the ansatz 
\begin{eqnarray}
B(r,\theta)=3B(r)C^{-1/2}_{l+1}(\cos\theta)
\label{sep-b}
\end{eqnarray}
and 
\begin{eqnarray}
Q(r,\theta)=Q(r)C^{-3/2}_{l+2}(\cos\theta),
\label{sep-q}
\end{eqnarray} 
where $C^{\nu}_{n}$ is the Gegenbauer function of order $n$ and index 
$\nu$ \cite{abram}. It turns out that the relations given in Ref. 
\cite{chandra} between the Gegenbauer functions and the Legendre functions 
are correct only for $l\ge 2$. (See Eqs. (21) and (22) in Chapter 4 of Ref. 
\cite{chandra}.) For $l=1$, however, they give erroneous results. The 
correct expression for the $n=3$ and $\nu=-3/2$ Gegenbauer function is 
$C^{-3/2}_{3}(\cos\theta)=\frac{1}{2}(\cos^{3}\theta-3\cos\theta)$,
whereas the expressions given in Ref. \cite{chandra} give (incorrectly) an 
identically-vanishing expression. For reference, we give here an expression 
from which the Gegenbauer functions can be calculated directly for all
orders and indices \cite{abram}:
\begin{eqnarray*}
C^{\nu}_{n}(\cos\theta)=\sum_{m=0}^{n}\frac{\Gamma(\nu+m)\Gamma
(\nu+n-m)}{m!(n-m)![\Gamma(\nu)]^{2}}\cos(n-2m)\theta.
\end{eqnarray*}

Substituting Eqs. (\ref{sep-b}) and (\ref{sep-q}) in Eqs. (\ref{eq1}) and 
(\ref{eq2}) we obtain the radial equations
\begin{eqnarray}
\Delta\frac{d}{\,dr}\left(\frac{\Delta}{r^{4}}\frac{\,dQ}{\,dr}\right)
&-&(l-1)(l+2)\frac{\Delta}{r^{4}}Q+\sigma^{2}Q\nonumber\\
&=&-\frac{4Q_{*}}{r^{3}}(l-1)(l+2)\Delta e^{\nu}B
\label{eq3}
\end{eqnarray}
and 
\begin{eqnarray}
\frac{d}{\,dr}\left[e^{2\nu}\frac{d}{\,dr}(re^{\nu}B)\right]-
l(l+1)\frac{e^{\nu}}{r}B\nonumber\\
+\left(\sigma^{2}re^{-\nu}-\frac{4Q_{*}^{2}}
{r^{3}}e^{\nu}\right)B
=-Q_{*}\frac{Q}{r^{4}}.
\label{eq4}
\end{eqnarray}

\section{Derivation of the coupled equations}
We now define the functions
\begin{eqnarray}
\bar{H}_{1}^{(-)}=-2re^{\nu}B(r)
\label{h1}
\end{eqnarray}
and
\begin{eqnarray}
\bar{H}_{2}^{(-)}=\frac{1}{r}Q(r).
\label{h2}
\end{eqnarray}
Note, that here we deviated from the formalism of Ref. \cite{chandra}. The 
reason for this deviation is as follows: In Eq. (142) of Chapter 5 of Ref. 
\cite{chandra} the variables $\mu$ and $n$ (do not confuse $n$ here with the 
order of the Gegenbauer functions)
are defined by $\mu^{2}=2n=(l-1)(l+2)$. Then, when the function 
$H_{1}^{(-)}$ is subsequently defined (Eq. (143) of Chapter 5 of Ref.
\cite{chandra}) it is divided by $\mu$. However, for dipole modes $l=1$, and 
consequently $\mu$ vanishes identically. As $B$ and $H_{1}^{(-)}$ are 
physically-meaningful functions [they determine the electromagnetic field via 
Eq. (\ref{b}) and the Maxwell equations] 
this is manifestly inappropriate. By keeping this 
ill-defined functions for the treatment of dipole perturbations, one might 
encounter divisions and multiplications of finite physical fields by 
identically-vanishing expressions, and thus obtain nonsensical results. It 
should be stressed, however, 
that for all other modes, namely, for $l\ge 2$, the 
perturbative formalism as presented in Ref. \cite{chandra} is perfectly 
correct and valid. 

We now change variables to the Regge-Wheeler ``tortoise'' coordinate 
$r_{*}$ defined by $d/\,dr_{*}=(\Delta/r^{2})d/\,dr$, and find that 
$\bar{H}_{1}^{(-)}$ and $\bar{H}_{2}^{(-)}$ satisfy a pair of coupled 
second-order differential equations 
\begin{eqnarray}
\Lambda^{2}\bar{H}_{1}^{(-)}&=&\frac{\Delta}{r^{5}}\left\{\left[l(l+1)r
-3M+4\frac{Q_{*}^2}{r}\right]\bar{H}_{1}^{(-)}\right.\nonumber\\
&+&\left.3M\bar{H}_{1}^{(-)}
+2Q_{*}\bar{H}_{2}^{(-)}\right\}
\label{eq5}
\end{eqnarray} 
and
\begin{eqnarray}
\Lambda^{2}\bar{H}_{2}^{(-)}&=&\frac{\Delta}{r^{5}}\left\{\left[l(l+1)r
-3M+4\frac{Q_{*}^2}{r}\right]\bar{H}_{2}^{(-)}\right.\nonumber\\
&-&\left. 3M\bar{H}_{2}^{(-)}+2Q_{*}(l-1)(l+2)\bar{H}_{1}^{(-)}\right\}.
\label{eq6}
\end{eqnarray}
Here, $\Lambda^{2}=d^{2}/\,dr_{*}^{2}+\sigma^{2}.$ 
It is important to notice that Eqs. (\ref{eq5}) and (\ref{eq6}) are already 
decoupled in the following two cases: First, when $Q_{*}=0$, namely, when 
the electric charge of the black hole vanishes and the 
Reissner-Nordstr\"{o}m black hole degenerates to Schwarzschild. In this 
case there should be just one radiative perturbative mode, described by 
the Regge-Wheeler equation, which describes the perturbations of
Schwarzschild. Indeed, we find that Eq. (\ref{eq6}) reduces to 
the Regge-Wheeler equation in the limit of vanishing electric charge. The 
other equation apparently describes another radiative mode (as it is a 
wave-like differential equation). We shall discuss the meaning of the
other apparent mode in the next section. 
Second, when $l=1$, namely, in the case of dipole perturbations. (In fact, 
in this case Eq. (\ref{eq6}) governing $\bar{H}_{2}^{(-)}$ is decoupled, 
while Eq. (\ref{eq5}) still couples $\bar{H}_{1}^{(-)}$ and 
$\bar{H}_{2}^{(-)}$. However, in such a case it is very easy to decouple 
the equations.) In the next section, after we decouple the equations, we 
shall discuss this case in detail.

\section{The decoupling of the equations}

\subsection{Decoupling the equations for the dipole case}

In a similar way to the decoupling procedure we used in the polar case in 
Paper I, we first examine the $l=1$ case. As already noted in 
the preceding section, in this case the equation governing the evolution 
of $\bar{H}_{2}^{(-)}$ is already decoupled from $\bar{H}_{1}^{(-)}$. It is 
easy to verify [by direct 
substitution in Eqs. (\ref{eq5}) and (\ref{eq6})] that 
the complete decoupling can be obtained by a transformation to the new 
radial functions $\bar{Z}_{1}^{(-)}$ and $\bar{Z}_{2}^{(-)}$ defined by 
\begin{eqnarray}
\bar{H}_{1}^{(-)}=\frac{1}{q_{1}}\left(\bar{Z}_{1}^{(-)}+2
\frac{Q_{*}}{q_{1}}\bar{Z}_{2}^{(-)}\right)
\label{dec1l=1}
\end{eqnarray}
and 
\begin{eqnarray}
\bar{H}_{2}^{(-)}=-\frac{6M}{q_{1}^{2}}\bar{Z}_{2}^{(-)}.
\label{dec2l=1}
\end{eqnarray}
Here, $q_{1}$ is an arbitrary parameter. (It will be fixed when we decouple 
the equations for any $l$ below.) The factor $(-6M)$ in Eq.
(\ref{dec2l=1}) was 
taken for convenience. We then find the equations for the functions 
$\bar{Z}_{1}^{(-)}$ and $\bar{Z}_{2}^{(-)}$ to be   
\begin{eqnarray}
\Lambda^{2}\bar{Z}_{1}^{(-)}=\frac{\Delta}{r^{5}}\left(2r+4\frac{Q_{*}^2}{r}
\right)\bar{Z}_{1}^{(-)}
\label{z1l=1}
\end{eqnarray}
and 
\begin{eqnarray}
\Lambda^{2}\bar{Z}_{2}^{(-)}=\frac{\Delta}{r^{5}}\left(2r-6M
+4\frac{Q_{*}^2}{r}
\right)\bar{Z}_{2}^{(-)}.
\label{z2l=1}
\end{eqnarray}
Eq. (\ref{z2l=1}) is just the Regge-Wheeler equation for the case $l=1$, 
and describes a true physical radiative mode (for the electromagnetic 
field). Eq.~(\ref{z1l=1}), however, does not describe any physical mode, 
as it is nothing but the evolution equation for the {\it violation} of the
momentum constraint, which propagates hyperbolically on its own
(cf. Eq.~(68) of Ref. \cite{anderson98} and the discussion therein). 
If we assume the momentum constraint to be satisfied on the initial
spacelike hypersurface 
(namely, if we take both $\bar{Z}_{1}^{(-)}$ and its time derivative to
vanish on the initial slice), then $\bar{Z}_{1}^{(-)}$ will vanish
forever, and consequently Eq. (\ref{z1l=1}) carries no physical
information. Consequently, both $\bar{H}_{1}^{(-)}$ and
$\bar{H}_{2}^{(-)}$ are proportional to $\bar{Z}_{2}^{(-)}$, and  
Eqs. (\ref{dec1l=1}) and (\ref{dec2l=1}) are diffeomorphic and describe
the same physical radiative mode of the electromagnetic field. The reason
for the appearance of the equation for the evolution of the violation of
the constraint equation is clear. The field equations we use are
unconstrained, and therefore one should not be surprised to find
non-trivial equations which describe the unphysical sector of the
unconstrained theory. The field equations we used are the $\phi \, r$ and
the $\phi \, \theta$ components of the Einstein equations. In order to
have a constrained theory, one should include also the $t\, \theta$ and $t
\, \phi$ components, which will insure that unphysical modes will not
appear. 

Hence, we conclude that in this case there is just one 
radiative mode, as expected from the non-radiative character of the 
dipole gravitational field. We add here that in the dipole
mode of the polar perturbations  
(see Paper I) all the physical observables (i.e., the metric 
coefficients and the components of the Maxwell tensor) are obtained from 
just one differential equation, even though there appear to be two such 
equations, again as should be expected from the nature of the dipole
modes. As in the axial case we study here, again the unphysical mode is an
autonomous evolution of the violation of the momentum constraint. 
In the axial case one should preclude the other differential 
equation for the evolution of $\bar{Z}_{1}^{(-)}$, and thus find that the 
functions $\bar{H}_{1}^{(-)}$ and $\bar{H}_{2}^{(-)}$ are diffeomorphic 
(they are distinguished by just a multiplicative factor).

\subsection{Decoupling the equation for the general case}

We now turn to the decoupling of the equations in the general case (i.e., 
for any $l$). We shall follow the decoupling procedure of Paper I. We seek
functions $\bar{Z}_{1}^{(-)}$ and 
$\bar{Z}_{2}^{(-)}$ given by the ansatz 
\begin{eqnarray}
\bar{H}_{1}^{(-)}=\alpha\bar{Z}_{1}^{(-)}+\beta\bar{Z}_{2}^{(-)}
\label{an1}
\end{eqnarray}
and 
\begin{eqnarray}
\bar{H}_{2}^{(-)}=\gamma\bar{Z}_{1}^{(-)}+\delta\bar{Z}_{2}^{(-)},
\label{an2}
\end{eqnarray} 
where $\alpha ,\beta ,\gamma$, and $\delta$ are parameters yet to be fixed. 
Substitution into Eqs. (\ref{eq5}) and (\ref{eq6}) yields 
\begin{eqnarray}
\alpha\Lambda^{2}\bar{Z}_{1}^{(-)}&+&\beta\Lambda^{2}\bar{Z}_{2}^{(-)}
\nonumber\\
&=& 
\frac{\Delta}{r^{5}}\left[2Q_{*}\gamma+l(l+1)r\alpha
+4\frac{Q_{*}^{2}}
{r}\alpha\right]\bar{Z}_{1}^{(-)}\nonumber\\
&+&\frac{\Delta}{r^{5}}\left[2Q_{*}\delta+l(l+1)r\beta+4\frac{Q_{*}^{2}}
{r}\beta\right]\bar{Z}_{2}^{(-)}
\label{eq7}
\end{eqnarray}
and
\begin{eqnarray}
\gamma\Lambda^{2}\bar{Z}_{1}^{(-)}&+&\delta\Lambda^{2}\bar{Z}_{2}^{(-)}
=\frac{\Delta}{r^{5}}\left[2Q_{*}(l-1)(l+2)\alpha\right.\nonumber\\
&+&l(l+1)r\gamma-6M\gamma+4\frac{Q_{*}^{2}}
{r}\gamma ]\bar{Z}_{1}^{(-)}\nonumber\\
&+&\frac{\Delta}{r^{5}}[2Q_{*}(l-1)(l+2)\beta+l(l+1)r\delta\nonumber\\
&-&6M\delta
+4\frac{Q_{*}^{2}}
{r}\delta]\bar{Z}_{1}^{(-)}.
\label{eq8}
\end{eqnarray}

We now multiply Eq. (\ref{eq7}) by $\gamma$ and Eq. (\ref{eq8}) by 
$\alpha$, and subtract the two equations. To have the equations decoupled we 
require that 
$$2Q_{*}\gamma^{2}+6M\alpha\gamma-2Q_{*}(l-1)(l+2)\alpha^{2}=0.$$ 
This is a quadratic equation in $\gamma$, say, whose solution can be 
determined uniquely, once we recall that $\gamma$ vanishes for $l=1$. 
Hence, we find that 
\begin{eqnarray*}
\gamma=-\frac{1}{2}\frac{\alpha}{Q_{*}}\left[3M-
\sqrt{9M^{2}+4Q_{*}^{2}(l-1)(l+2)}\right],
\end{eqnarray*}
or, 
\begin{eqnarray}
\gamma=-\frac{1}{2}\frac{\alpha}{Q_{*}}q_{2}, 
\end{eqnarray}
where $q_{2}\equiv 3M-
\sqrt{9M^{2}+4Q_{*}^{2}(l-1)(l+2)}$. 

Next we multiply Eq. (\ref{eq7}) by $\delta$ and Eq. (\ref{eq8}) 
by $\beta$, and subtract the equations. To have the equations decoupled we now 
require that 
$$2Q_{*}\delta^{2}+6M\beta\delta-2Q_{*}(l-1)(l+2)\beta^{2}=0.$$ 
Solving this equation for $\delta$, and requiring that $\delta$ will not 
vanish for $l=1$, we find that 
\begin{eqnarray*}
\delta=-\frac{1}{2}\frac{\beta}{Q_{*}}\left[3M+
\sqrt{9M^{2}+4Q_{*}^{2}(l-1)(l+2)}\right],
\end{eqnarray*} 
or, 
\begin{eqnarray}
\delta=-\frac{1}{2}\frac{\beta}{Q_{*}}q_{1},
\end{eqnarray}
where $q_{1}\equiv 3M+
\sqrt{9M^{2}+4Q_{*}^{2}(l-1)(l+2)}$. The parameters $\alpha$ and $\beta$
remain to be fixed arbitrarily, although one could let them equal their
dipole-case counterparts. 

Substitution of the expressions for the decoupling parameters and the 
ansatz (\ref{an1}) and (\ref{an2}) in Eqs. (\ref{eq5}) and (\ref{eq6}) we 
find that the two decoupled wave equations assume the form
\begin{eqnarray}
\Lambda^{2}\bar{Z}_{1}^{(-)}&=&\frac{\Delta}{r^{5}}\left[l(l+1)r
+4\frac{Q_{*}^{2}}{r}\right.\nonumber \\
&& \left. +\frac{q_2}{q_2-q_1}\left(2q_1-6M\right)\right]
\bar{Z}_{1}^{(-)},
\end{eqnarray}
and
\begin{eqnarray}
\Lambda^{2}\bar{Z}_{2}^{(-)}&=&\frac{\Delta}{r^{5}}\left[l(l+1)r
+4\frac{Q_{*}^{2}}{r}\right.\nonumber \\
&& \left. +\frac{q_1}{q_1-q_2}\left(2q_2-6M\right)\right]
\bar{Z}_{2}^{(-)}.
\end{eqnarray}
These equations can be put in a more compact and symmetrical form as 
\begin{eqnarray}
\Lambda^{2}\bar{Z}_{i}^{(-)}=V_{i}^{(-)}\bar{Z}_{i}^{(-)},
\end{eqnarray}
where the effective potential is given by 
\begin{eqnarray}
V_{i}^{(-)}=\frac{\Delta}{r^{5}}\left[l(l+1)r+4\frac{Q_{*}^{2}}{r}
-q_j\right],
\end{eqnarray}
and $i,j=1,2$, $\;i\ne j$. This wave equation is similar to the equation
given in Ref. \cite{chandra}. However, in our case this equation holds for
all the multipole modes, including the dipole, whereas in Ref.
\cite{chandra} the dipole modes are excluded, because of the
inapplicability of the perturbative formalism to the dipole modes. 

\section{The completion of the solution}
We can now complete the solution be giving the metric coefficients and the 
components of the Maxwell tensor in terms of the functions
$\bar{Z}_{i}^{(-)}$ (for dipole modes just   
$\bar{Z}_{2}^{(-)}$). Knowing the functions $\bar{Z}_{i}^{(-)}$ the 
functions $\bar{H}_{i}^{(-)}$ can be calculated using Eqs. (\ref{an1}) 
and (\ref{an2}). Then, the functions $Q$ and $B$ can be computed by Eqs. 
(\ref{h1}) and (\ref{h2}). Using Eq. (\ref{b}) and the (linearized) Maxwell 
equations (\ref{max1}), (\ref{max2}), and (\ref{max3}) all the perturbed 
components of the Maxwell tensor are calculable. From Eq. (\ref{h2}) the 
function $Q$ is known. Then, due to Eq. (\ref{q}), we know the combination 
$q_{2,3}-q_{3,2}$ of the perturbations of the 
metric coefficients. Using Eqs. (\ref{dif1}) and 
(\ref{dif2}) we can now obtain explicit expressions for the metric 
coefficients, and thus complete the solution.

We now have a perturbative formalism which is valid for the treatment of 
all the multipole modes, including the dipole, 
and which generalizes the formalism given 
in Ref. \cite{chandra}. (The formalism for the polar modes is given in
Paper I.) In a similar way to Ref. \cite{chandra}, one can 
now show the relations between the polar and the axial modes. (This is done 
most easily via the Newman-Penrose formalism---see Ref. \cite{chandra}.) 

\section*{Acknowledgments}
I thank 
Amos Ori for numerous invaluable discussions and for useful comments.

\end{document}